\documentclass[a4paper,11pt]{article}
\pdfoutput=1 % if your are submitting a pdflatex (i.e. if you have
\usepackage{jheppub,bm} 

\usepackage{tikz}
\usepackage{pgfplots}
\pgfplotsset{compat=1.14}
\usepackage{tikz-3dplot}
\usetikzlibrary{intersections, calc,positioning,decorations.pathreplacing,decorations.pathmorphing,arrows,arrows.meta,patterns,pgfplots.fillbetween}

\usepackage[T1]{fontenc} % if needed
\usepackage[utf8]{inputenc}

\newcommand{\dd}{\mathrm{d}}

\def\CB{\mathcal{B}}

\def\CD{\mathcal{D}}

\def\CJ{\mathcal{J}}

\def\CM{\mathcal{M}}

\def\CQ{\mathcal{Q}}

\title{Does Boundary Distinguish Complexities?}
\author[a]{Yoshiki Sato}
\author[b]{and Kento Watanabe}

\affiliation[a]{Kavli Institute for the Physics and Mathematics of the Universe (WPI),\\ 
The University of Tokyo Institutes for Advanced Study, The University of Tokyo,\\
Kashiwa, Chiba 277-8583, Japan}
\affiliation[b]{Department of Physics, Faculty of Science,
The University of Tokyo,\\
Bunkyo-ku, Tokyo 113-0033, Japan}

\abstract{Recently, Chapman \textit{et al}. argued that holographic complexities for defects distinguish action from volume. 
Motivated by their work, we study complexity of quantum states in conformal field theory with boundary. 
In generic two-dimensional BCFT, we work on the path-integral optimization which gives one of field-theoretic definitions for the complexity.
We also perform holographic computations of the complexity in Takayanagi's AdS/BCFT model following by the ``complexity $=$ volume'' conjecture and ``complexity $=$ action'' conjecture.
We find that increments of the complexity due to the boundary show the same divergent structures in these models except for the CA complexity in the AdS$_3$/BCFT$_2$ model as the argument by Chapman \textit{et al}. Thus, we conclude that boundary does not distinguish the complexities in general.}

\preprint{UT-19-20, IPMU19-0118}

\begin{document} 
\maketitle
\flushbottom

\section{Introduction}

It is widely accepted that the idea of quantum information is useful and important to understand quantum gravity in the AdS/CFT correspondence, starting from the work \cite{Ryu:2006bv}.
The holographic entanglement entropy \cite{Ryu:2006bv} is given by an extremal area of a codimension-two surface anchored to the entangling surface on the AdS boundary.
There are many significant developments based on the holographic entanglement entropy (See \cite{Rangamani:2016dms} for review and references therein). 
However, the holographic entanglement entropy is not enough to understand a black hole physics, for instance, a late time dynamics of black holes \cite{Susskind:2014moa}.

To remedy this problem, Susskind and his collaborators introduced a notion of complexity of quantum states in the context of the AdS/CFT correspondence and proposed its holographic dual.
Now there are two different conjectures for the complexity; the ``complexity $=$ volume'' (CV) conjecture  \cite{Susskind:2014,Susskind:2014rva} and the ``complexity $=$ action'' (CA) conjecture \cite{Brown:2015bva,Brown:2015lvg}.
The CV conjecture states that the holographic dual of the complexity is given by a maximal volume of codimension-one surface anchored to the AdS boundary\footnote{Alternatively quantum information metric has been studied as a field-theoretic dual of the maximal volume  \cite{MIyaji:2015mia,Belin:2018fxe,Belin:2018bpg}.},
\begin{align}
 C_{\text{V}} = \frac{V}{G_{\text{N}}L} \label{CVcon} \,,
\end{align}
where $G_{\text{N}}$ is the Newton constant.
In order to make the complexity dimensionless, the length scale $L$ is introduced and this is assumed to be an AdS radius.
The CV conjecture is ambiguous due to the length scale.
On the other hand, the CA conjecture states that the holographic dual of the complexity is given by the gravitational action on the Wheeler-DeWitt (WDW) patch,
\begin{align}
 C_{\text{A}} = \frac{I_{\text{WDW}}}{\pi \hbar} \,,  \label{CAcon}
\end{align}
where $\hbar$ is the Planck constant and we will set $\hbar =1$ later. In the CA conjecture, there is no length scale introduced by hand.
It has been revealed that there is no significant difference between the two conjectures such as they show the same divergence structures, late time behaviours and so on.

The CV and CA conjectures were originally proposed as holographic duals of the complexity but its  definition in quantum field theory did not exist at that time.
Now there are some proposals of the definition of the complexity in quantum field theory. 
One of the authors and his collaborators proposed the complexity by optimizing the path integral appearing in the wave functional \cite{Caputa:2017urj,Caputa:2017yrh}, called the path-integral optimization we will work in this paper.
One of the advantages of the path-integral optimization is that we can obtain the complexity in generic CFT. 
Furthermore, the path-integral optimization can be regarded as a special case of the circuit complexity \cite{Jefferson:2017sdb} shown by \cite{Camargo:2019isp}. See \cite{Bhattacharyya:2018wym,Takayanagi:2018pml} for a further development of the path-integral optimization.
Note that these definitions of the complexity are not satisfactory  because they contain some ambiguities 
\cite{Jefferson:2017sdb,Chapman:2017rqy,Caputa:2017urj,Caputa:2017yrh,Caputa:2018kdj}.

Recently, the authors in \cite{Chapman:2018bqj} argued that defects might distinguish the features of these two holographic conjectures. 
They showed that the increments of the holographic complexities in a AdS$_3$/CFT$_2$ model with a defect \cite{Azeyanagi:2007qj} behave  
\begin{align}
 \Delta C_{\text{V}}^{\text{defect}} =  C_{\text{V}}^{\mathrm{DCFT}} -  C_{\text{V}}^{\mathrm{CFT}} \neq 0 \,, 
 \qquad \Delta C_{\text{A}}^{\text{defect}} =   C_{\text{A}}^{\mathrm{DCFT}} -  C_{\text{A}}^{\mathrm{CFT}} = 0 \,.
\end{align}
It implies that the defects are detected by $C_{\text{V}}$ but invisible to $C_{\text{A}}$.
They also showed that the circuit complexity \cite{Jefferson:2017sdb} of several models in defect CFT$_2$ does not depend on the presence of defects.
Hence, their result suggests that the CV conjecture is not adequate for the holographic dual of the complexity.

In this paper, we will make an attempt to test their argument in boundary CFT (BCFT). 
Because two copies of BCFT can be regarded as a CFT with a codimension-one defect or interface via doubling trick, 
 our setup is relevant to defect CFTs considered in \cite{Chapman:2018bqj}. 
We will compute some quantities conjectured to be dual to the complexity of quantum states. 
One is the optimized Liouville action $C_{\text{L}}$ in the path-integral optimization approach \cite{Caputa:2017urj,Caputa:2017yrh} in BCFT$_2$.
The others are the maximal volume $C_{\text{V}}$ and the WdW action $C_{\text{A}}$ in a holographic model proposed by Takayanagi \cite{Takayanagi:2011zk,Fujita:2011fp}.
Especially we will study the boundary complexity which is an increment of the complexity due to the presence of the boundary,
\begin{align}
 \Delta C^{\text{bdy}} =  C^{\mathrm{BCFT}} - \frac{1}{2} C^{\mathrm{CFT}} \,. \label{bdyC}
\end{align}
The factor $1/2$ comes from the fact that the spacetime of BCFT is just half of the spacetime of CFT.
We will check whether the boundary complexities in these three proposals depend on the existence of the boundary.

The organization of this paper is as follows.
In the next section, we apply the path-integral optimization to BCFT$_2$ and compute a boundary complexity.
We will see that the path-integral optimization naturally produces a geometry in Takayanagi's AdS$_3$/BCFT$_2$ model.
Hence, in order to compare the result in section \ref{2}, we give a brief review of the AdS$_{d+1}$/BCFT$_d$ model and study holographic complexities following the CV and CA conjectures in section \ref{3}.
The final section is devoted to discussion.

\paragraph{Note added:} After submitting the paper to arXiv, we were informed of a forthcoming paper by P.~Braccia, A.~Cotrone and E.~Tonni \cite{BCT}, which is based on a thesis written by P.~Braccia presented in the end of July and we received the thesis where the correct WDW patch in the AdS$_3$/BCFT$_2$ setup has already been presented. 
They independently studied the CV and CA complexities in the AdS$_3$/BCFT$_2$ setup with finite interval and there are some overlaps with section \ref{3} in our paper.

%%%%%%%%%%%%%%%%%%%%%%%%%%%%%%%%%%%%

\section{Path-integral optimization in BCFT}
\label{2}

In this section, we will work on the path-integral optimization in BCFT to compute the optimized Liouville action proposed as the complexity of the ground state \cite{Caputa:2017urj,Caputa:2017yrh}.
To simplify our discussion, we restrict our attention to BCFT$_2$ in this section.

\subsection{Ground state wave functional in BCFT and boundary Liouville action}

Consider a two-dimensional CFT on half line with a flat Euclidean metric,
\begin{align}
    \dd s^2=\delta_{ab}\dd x^a \dd x^b =\dd z^2 + \dd x^2 \,.
\end{align}
The ground state wave functional is described by a Euclidean path integral on a two-dimensional region $\mathcal{M}= \{x \geq 0, \epsilon \leq z < \infty \}$
\begin{align}
 \Psi_{\delta_{ab}}^{\mathrm{BCFT}} [\tilde{\varphi}(x)] 
 = \int_\CM \! \CD \varphi \, \mathrm{e}^{-S_{\mathrm{BCFT}}[\varphi]} \,  \prod_{x>0} \delta (\varphi(\epsilon, x) - \tilde{\varphi}(x)) \,, 
\end{align}
where $\tilde{\varphi} (x)$ is a configuration of the CFT field at the cutoff surface $\partial \mathcal{M}_0 = \{x \geq 0, z = \epsilon \}$. 
The boundary condition on the other boundary $\partial \mathcal{M}_1 = \{ x = 0, \epsilon \leq z < \infty \}$ classifies types of BCFTs. 
We introduce a cutoff parameter $\epsilon$ for later convenience.
See the left picture in Figure \ref{fig:opt1}.

For the purpose to estimate the wave functional effectively, the path integral is actually redundant because some high-energy degrees of freedom would be suppressed in the deep region of the bulk $\mathcal{M}$. 
To reduce such degrees of freedom, 
we deform the background metric with a boundary condition keeping the wave functional.
In two-dimensional CFTs, it can be realized by Weyl transformation of the background metric,  
\begin{align}
 \delta_{ab} \ \to \ \mathrm{e}^{2\phi} \delta_{a b} \,.
\end{align}
This procedure is analogous to a coarse-graining procedure for discretized path-integral on a flat lattice with spacing $\epsilon$ 
due to deforming the flat lattice to a lattice with position dependent spacing $\epsilon \mathrm{e}^{-\phi}$, as firstly considered in \cite{Miyaji:2016mxg}.

Under the Weyl transformation of the reference metric $\delta_{ab}$, 
the wave functional transforms as  
\begin{align} \label{wave}
 \Psi_{\mathrm{e}^{2\phi} \delta_{a b}}^{\text{BCFT}} [\tilde{\varphi}(x)] 
 = \mathrm{e}^{S_{\text{L}}[\phi]-S_{\text{L}}[0]} \, \Psi_{\delta_{ab}}^{\text{BCFT}} [\tilde{\varphi}(x)] \,.
\end{align}
Here $S_{\text{L}}$ is the boundary Liouville action\footnote{Here we rescale $\phi$, $\mu$ and $\mu_{\mathrm{B}}$ in \cite{Fateev:2000ik} as $b \phi \to \phi$, $4\pi b^2 \mu \to \mu$ and $2\pi b^2 \mu_{\mathrm{B}} \to \mu_{\mathrm{B}}$, respectively, where $b$ is a coupling relevant to the central charge $c = 1 + 6 (b + 1/b)^2$ ($c\simeq 6/b^2$ in the semi-classical limit $b \to 0$). By applying the rescaling to (3.5) in \cite{Fateev:2000ik} associated with boundary two-point functions, we can find the quantum constraint for the existence of the semi-classical solutions, $\mu_{\mathrm{B}}^2/\mu = \pi b^2/(2\tan (\pi b^2/2))$. The range of the cosmological constant is $0 \leq |\mu_{\mathrm{B}}/\sqrt{\mu}| \leq 1$ for $|b| \leq 1$. } \cite{Fateev:2000ik},
\begin{align} \label{LA}
\begin{aligned}
S_{\mathrm{L}} [\phi] &= \frac{c}{24 \pi} \int_{\mathcal{M}} \! \dd ^2x \, \sqrt{g} 
\left( R \phi + (\partial \phi)^2 + \mu \mathrm{e}^{2 \phi} \right) \\
&\quad + \frac{c}{12 \pi} \sum_{i} \int_{\partial \mathcal{M}_i} \! \dd s \, \sqrt{h} 
\left( K \phi + \mu_{\mathrm{B}}^{(i)} \mathrm{e}^{\phi} \right) \,,
\end{aligned}
\end{align}
with the central charge $c$, the metric $g_{ab}$, the Ricci scalar $R$, the induced metric on the boundary $h$ and the extrinsic curvature $K$.
The Liouville action is evaluated on the original metric $\delta_{ab}$ in \eqref{wave}. 
The parameters $\mu$ and $\mu_{\text{B}}^{(i)}$ represent the bulk and the boundary cosmological constants respectively. 
We will set $\mu_{\text{B}}^{(0)} = 0$, $\mu_{\text{B}}^{(1)} = \mu_{\text{B}}$ for later convenience,  
and take $\mu, \mu_{\mathrm{B}} \geq 0$ for the convergence of the action in the semi-classical level.
The appearance of the Liouville action in \eqref{wave} follows from a transformation of the path-integral measure, 
\begin{align}
 [\CD \varphi]_{\mathrm{e}^{2\phi}\delta_{ab}} 
 =  \mathrm{e}^{S_{\text{L}}[\phi]-S_{\text{L}}[0]} \, [\CD \varphi]_{\delta_{ab}} \,.
\end{align}
The overall factor reflects how much redundant degrees of freedom (or lattice sites) can be reduced. 
To optimize the path integral, we will minimize this factor, or the exponent $S_{\mathrm{L}}$. 
From the solution of the equation of motion for $\phi$, we will obtain the optimized path-integral geometry. 
Then the on-shell Liouville action is expected to be a measure for complexity of quantum states in CFTs.

\subsection{Optimize the Liouville action}

Let us move to the analysis of the Liuouville action \eqref{LA}.
The action leads the equation of motion and the boundary condition,
\begin{align}
 - 2\partial^2 \phi + 2\mu \mathrm{e}^{2 \phi} &= 0 \,, \label{EOM} \\
 n \cdot \partial \phi + \mu_{\mathrm{B}}^{(i)} \mathrm{e}^{\phi} &= 0 \,. \label{BC1}
\end{align}
where $n^a$ is an out-going unit normal vector.
Note that the Ricci scalar and the extrinsic curvature vanish because the original metric is flat and the boundaries are conformal.
From the transformation laws under the Weyl transformation, 
\eqref{EOM} and \eqref{BC1} are simply written as 
\begin{align}
&R + 2 \mu = 0 \,, \\
&K + \mu_{\mathrm{B}}^{(i)} = 0 \,,
\end{align}
in the deformed background.
The geometry is the AdS spacetime with the AdS radius $L = 1/\sqrt{\mu}$.
It also has another length scale $\mu_{\text{B}} = \mu_{\text{B}}^{(1)}$ associated to the boundary $\partial \mathcal{M}_{1}$. 
We can make a dimensionless parameter $\mu_{\mathrm{B}} /\sqrt{\mu} = \mu_{\mathrm{B}} L$.

\begin{figure}
\centering
\begin{minipage}{0.3\hsize}
\begin{tikzpicture}[scale=1.30]
   \draw[-{Triangle[angle'=60,scale=0.8]}] (0,0)--(2,0) node [right] {$x$};
   \draw[-{Triangle[angle'=60,scale=0.8]}] (0,0)--(0,2) node [above] {$z$};
   \draw[red,thick] (0,0)--(0,1.8);
   \draw (1.2,0) node [below] {$\partial \mathcal{M}_{0}$};
   \draw (-0.5,0.8) node [above] {$\partial \mathcal{M}_{1}$};
   \draw[-{Triangle[angle'=60,scale=0.8]}] (-0.4,0.9) arc (180:270:0.4);
   \draw (0.2,0)--(0.2,0.2)--(0,0.2);
   \filldraw[pattern=north east lines, pattern color=gray,opacity=0.6,draw=none] (0,0)--(0,2)--(-1,2)--(-1,0)--cycle;
   \draw (1,1) node {$\mathcal{M}$};
\end{tikzpicture}
\end{minipage}
\begin{minipage}{0.2\hsize}
\begin{tikzpicture}
   \draw[-{Triangle[angle'=60,scale=0.8]}] (0,1.2)--(1,1.2) node [above] {Optimize}--(2,1.2);
\end{tikzpicture}
\end{minipage}
\begin{minipage}{0.3\hsize}
\begin{tikzpicture}[scale=1.30]
   \draw[-{Triangle[angle'=60,scale=0.8]}] (0,0)--(2,0) node [right] {$x$};
   \draw[-{Triangle[angle'=60,scale=0.8]}] (0,0)--(0,2) node [above] {$z$};
   \draw[red,thick] (0,0)--(-1,1.732);
   \draw (-1.2,1.732) node [above] {$x = -\alpha z$};
   \draw (1.2,0) node [below] {$\partial \mathcal{M}_{0}$};
   \draw (-1.2,0.8) node [above] {$\partial \mathcal{M}_{1}$};
   \draw[-{Triangle[angle'=60,scale=0.8]}] (-1,0.9) arc (210:300:0.4);
   \draw (0.3,0) arc (0:120:0.3);
   \filldraw[pattern=north east lines, pattern color=gray,opacity=0.6,draw=none] (0,0)--(-1,1.732)--(-1,0)--cycle;
   \draw (1.2,1) node {$\mathcal{M}$};
\end{tikzpicture}
\end{minipage}
\caption{(Left) The setup of the path integral for vacuum wave functional in BCFT. The boundary $\partial \mathcal{M}_{1}$ is located at $x=0$ and the state is realized at $\partial \mathcal{M}_{0} = \{x>0, z = \epsilon \sim 0 \}$. 
(Right) The setup after the path-integral optimization. The boundary $\partial \mathcal{M}_{1}$ is tilted.}
\label{fig:opt1}
\end{figure}
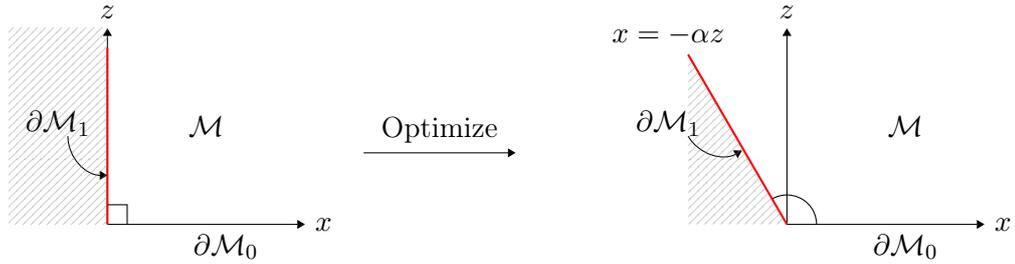

Taking the boundary condition such that the conformal factor decays at infinity and is fixed on the cutoff surface  $\partial \mathcal{M}_0$ as $\mathrm{e}^{2 \phi (z=\epsilon,x)}=L^2/\epsilon^2$, 
the equation of motion \eqref{EOM} leads
\begin{align}
\mathrm{e}^{2 \phi (z,x)} = \frac{L^2}{z^2} \,. \label{Lsol1}
\end{align}
Then the path-integral optimization leads the time slice of the AdS metric in Poincar\'{e} coordinates,
\begin{align}
\dd s ^2= L^2\frac{\dd z^2 + \dd x^2}{z^2}  \,. 
\end{align}
For the boundary $\partial \mathcal{M}_0$, we fix the shape by setting $\mu_{\mathrm{B}}^{(0)} = 0$ to keep the same wave functional. 
After the path-integral optimization, the boundary $\partial \mathcal{M}_0$ can be understood as the boundary of the AdS spacetime.
For the other boundary $\partial \mathcal{M}_1$ with $\mu_{\mathrm{B}} = \mu_{\mathrm{B}}^{(1)} \neq 0$, the boundary condition \eqref{BC1} determines the shape as\footnote{Note that, in order to have positive real $\alpha$, $\mu_{\text{B}} L$ is restricted to a specific region $0 \leq \mu_{\text{B}} L \leq 1$. This is consistent with the constraint from the quantum Liouville theory  mentioned in the previous footnote.}
\begin{align}
x = f(z) = -\alpha z \,, \qquad
\alpha = \frac{\mu_{\mathrm{B}}L}{\sqrt{1- \mu_{\mathrm{B}}^2L^2}} \,. \label{m1shape}
\end{align}
Hence, the path-integral optimization to BCFT introduces a new boundary in the radial direction of the AdS spacetime as Takayanagi's AdS/BCFT model \cite{Takayanagi:2011zk,Fujita:2011fp}.
For $\mu_{\mathrm{B}} L\to 0$ ($\alpha \to 0$) limit, $\partial \mathcal{M}_1$ becomes perpendicular to $\partial \mathcal{M}_{0}$ and no shape deformation happens after the optimization.  
For $\mu_{\mathrm{B}} L\to 1$ ($\alpha \to \infty$) limit, the corner between $\partial \mathcal{M}_{0}$ and $\partial \mathcal{M}_{1}$ disappears and $\mathcal{M}$ becomes the upper-half plane.
As seen later, the slope $\alpha$ is related with the boundary entropy.
For the brief picture of our optimization procedure, see Figure \ref{fig:opt1}.

Finally, we obtain the on-shell Liouville action, 
\begin{align}
C_{\mathrm{L}}^{\mathrm{BCFT}} 
&= \left. S_{\mathrm{L}} \right|_{\text{on-shell}} \\
&= \frac{c}{12 \pi} \cdot \frac{x_{\infty}}{\epsilon} +\frac{c}{6 \pi}\alpha \log \left( \frac{z_{\infty}}{\epsilon}\right) \,, \label{Lon1}
\end{align}
where $z_\infty$ and $x_\infty$ correspond to IR cutoffs.
The first term has volume divergence and is half of the Liouville action in CFT without boundary. 
The data of the boundary can be read off from the second term. 
In total, the path-integral optimization leads the boundary complexity,
\begin{align}
 \Delta C_{\mathrm{L}}^{\text{bdy}}
 = C_{\mathrm{L}}^{\mathrm{BCFT}} - \frac{1}{2} C_{\mathrm{L}}^{\mathrm{CFT}} 
 =  \frac{c}{6 \pi}\alpha \log \left( \frac{z_{\infty}}{\epsilon}\right) \,. \label{bdyCL}
\end{align}
The boundary complexity diverges logarithmically and this behaviour is consistent with the defect complexity in the CV conjecture \cite{Flory:2017ftd,Chapman:2018bqj}.

Note that the complexity can depend on the regularization scheme as similar to the entanglement entropy. 
Indeed, for the first term in \eqref{Lon1} which is volume divergent, 
we can see how they affect. 
The coefficient of the logarithmic term in \eqref{Lon1}, however, is independent of the regularization schemes,
\begin{align}
\left. \Delta C_{\mathrm{L}}^{\text{bdy}} \right|_{\mathrm{univ}} 
= \frac{c}{6\pi} \alpha 
= \frac{c}{6\pi} \frac{\mu_{\mathrm{B}} L}{\sqrt{1- \mu_{\mathrm{B}}^2 L^2}} \,, \label{Cuniv}
\end{align}
and is often called the universal term in this sense.

%%%%%%%%%%%%%%%%%%%%%%%%%%%%%%%%%%
\subsection{Boundary entropy}

In this subsection, 
we will compute the boundary entropy \cite{Affleck:1991tk,Friedan:2003yc}, which is an increment of the entanglement entropy due to the existence of the boundary, in our setup by using the path-integral optimization \cite{Caputa:2017urj,Caputa:2017yrh}. 
From the result, we will find a parameter matching between the boundary entropy in BCFT and $\mu_{\mathrm{B}}L$ in the Liouville side. 

\begin{figure}
\centering
\begin{tikzpicture}[scale=1.50]
   \draw[-{Triangle[angle'=60,scale=0.8]}] (0,0)--(2.5,0) node [right] {$x$};
   \draw[-{Triangle[angle'=60,scale=0.8]}] (0,0)--(0,2) node [above] {$z$};
   \draw[red,thick] (0,0)--(-1,1.732);
   \draw (-1.2,1.732) node [above] {$x = -\alpha z$};
   \draw (0.7,-0.05) node [below] {$\partial \mathcal{M}_{A}$};
   \draw (2,-0.05) node [below] {$\partial \mathcal{M}_{\bar{A}}$};
   \draw (-0.65,0.3) node [above] {$\partial \mathcal{M}_{1}$};
   \draw[thick] (1.4,0) arc (0:120:1.4);
   \fill (1.4,0) circle (1.2pt) node[below] {$l$} (-0.7,1.732*0.7) circle (1.2pt);
   \filldraw[pattern=north east lines, pattern color=gray,opacity=0.6,draw=none] (0,0)--(-1,1.732)--(-1,0)--cycle;
   \draw[blue] (0,0)--(0.7,0) node [above] {$A$}--(1.4,0);
   \draw (0,0) node [below] {$0$};
   \draw (1.1,1.1) node [above] {$\gamma_{A}$};
\end{tikzpicture}
\caption{The entanglement entropy associated to the subsystem $A$ ($0 \leq x \leq l$) is given by the length of the arc $\gamma_{A}$ anchored on the boundary surface $\partial \mathcal{M}_{1}$ ($x = - \alpha z$) and the edge of $A$.}
\label{fig:bdyEE}
\end{figure}
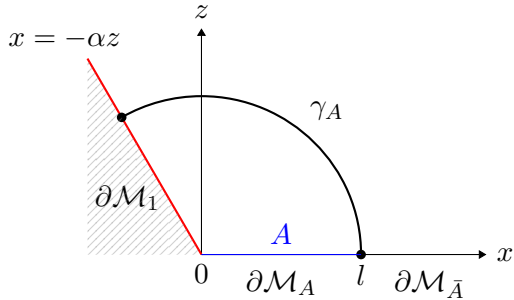

Consider a subsystem $A$ in the half line $x \geq 0$ as shown in Figure \ref{fig:bdyEE}. It has length $l$ and is attached to the boundary $x=0$. 
To compute the boundary entropy \cite{Affleck:1991tk,Friedan:2003yc} 
\begin{align} \label{defbdyEE}
     S_{\text{bdy}} = S_A^{\text{BCFT}} - \frac{1}{2} S_A^{\text{CFT}} \,, 
\end{align}
via the replica trick, we put a vertex operator at the edge of $A$ creating the deficit angle $2\pi(1-n)$.
Then the delta functional source term is added to the Liouville action and the equation of motion is deformed as 
\begin{align}
-\partial^2 \phi + \mu \mathrm{e}^{2 \phi} =  \pi (n-1) \cdot  \delta(x-l) \delta(z=0) \,,
\end{align}
with the boundary condition \eqref{BC1}.

In the path-integral optimization procedure for the reduced density matrix $\rho_{A}$, 
we divide the boundary $ \partial \mathcal{M}_{0}$ as 
$ \partial \mathcal{M}_{0} = \partial \mathcal{M}_{A} \cup \partial \mathcal{M}_{\bar{A}}$ associated to the subsystem $A$ and the compliment $\bar{A}$. 
We fix $\partial \mathcal{M}_{A}$ and deform $\partial \mathcal{M}_{\bar{A}}$  with $\mu_{\mathrm{B}} = \pi (1-n)$ so that the $n$-sheeted replica manifold $\mathcal{M}^{(n)}$ is realized. 
For $\partial \mathcal{M}_{1}$, we put the same boundary condition as before. 
To obtain the entanglement entropy, 
we take the limit $n \to 1$. 
In this limit, back reactions from the bulk vertex operator is suppressed and two boundaries $\partial \mathcal{M}_{1}$ and $\partial \mathcal{M}_{\bar{A}}$ are deformed independently.

We have the wave functional optimized by the path integral and, following previous works  \cite{Caputa:2017urj,Caputa:2017yrh}, can compute the entanglement entropy using it, 
\begin{align}
 S_{A}^{\text{BCFT}} 
 &= -\left. \partial_n \left( \log \frac{\mathrm{Tr} (\rho_{A}^n )}{(\mathrm{Tr} \rho_{A} )^n} \right) \right|_{n = 1} \\
 &=  \left. \partial_n \left( \frac{c (n-1)}{6} \int_{\gamma_A } \! \dd s \,  \mathrm{e}^{\phi} \right) \right|_{n = 1} \\
 &= \frac{c}{6} \log \left(\frac{2l}{\epsilon} \right) + \frac{c}{12} \log  \left(\frac{1 + \mu_{\mathrm{B}}L}{1 - \mu_{\mathrm{B}}L} \right) \,.
\end{align}
where $\gamma_{A}$ is an arc to which $\partial \mathcal{M}_{\bar{A}}$ transforms by the optimization and it is anchored on $\partial \mathcal{M}_{1}$ and the edge of $A$ (Figure \ref{fig:bdyEE}). 
The first term is half of the entanglement entropy in CFT without boundary and hence the second term represents the boundary entropy \eqref{defbdyEE},  
\begin{align}
 S_{\text{bdy}} = \frac{c}{12} \log \left(\frac{1 + \mu_{\mathrm{B}}L}{1 - \mu_{\mathrm{B}}L} \right) = \frac{c}{6} \, \mathrm{arcsinh} \, \alpha \,. \label{bdyEE}
\end{align}

Since the path-integral optimization naturally produces the AdS geometry with the cutoff in the radial direction, it is instructive to compare the boundary entropy we obtained with that of the AdS/BCFT model \cite{Takayanagi:2011zk}.
For this purpose, we introduce new coordinates, 
\begin{align}
    z=\frac{w}{\cosh (r/L)} \,, \quad x= w \tanh \left(\frac{r}{L}\right) \,.
\end{align}
In the new coordinates, the range of $w$ is $0\leq w < \infty$ and that of $r$ is $-r_\ast < r < \infty$. 
$-r_\ast$ is the position of the boundary in the radial direction and it is given by
\begin{align}
    r_\ast=\frac{L}{2}\log \left( \frac{1+\mu_{\mathrm{B}}L}{1-\mu_{\mathrm{B}}L} \right) \,. 
\end{align}
By using the Ryu-Takayanagi formula, the boundary entropy becomes $S_{\text{bdy}} =c r_\ast/6 L=r_\ast /4G_{\text{N}}$ with $c=3L/2G_{\text{N}}$ and it perfectly agrees with that of \cite{Takayanagi:2011zk}. Note that the range of the radial direction $r$ is different from $\rho$ used in \cite{Takayanagi:2011zk} but the relation $r_\ast = \rho_0$ holds. 

Finally, we can check the relation between the boundary entropy \eqref{bdyEE} and the universal coefficient of the boundary complexity \eqref{Cuniv}, 
\begin{align}
 \left. \Delta C_{\mathrm{L}}^{\text{bdy}} \right|_{\mathrm{univ}} 
 = \frac{c}{6 \pi} \sinh \left( \frac{6 S_{\mathrm{bdy}}}{c} \right) \,.
\end{align}
From this relation, the monotonicity of the boundary entropy under the renormalization group flow localized on the boundary, called the $g$-theorem \cite{Affleck:1991tk,Friedan:2003yc,Casini:2016fgb}, implies the monotonicity of the boundary complexity under the boundary RG flow.

%%%%%%%%%%%%%%%%%%%%%%%%%%%%%%%%%%%%
\section{Holographic complexity in the AdS/BCFT model}
\label{3}

In this section, we consider the holographic complexities $C_{\text{V}}$ and $C_{\text{A}}$ in the AdS/BCFT model proposed by Takayanagi \cite{Takayanagi:2011zk,Fujita:2011fp}.

\subsection{Review of the AdS/BCFT model}

Consider BCFT which is defined on a half plane, $x_1 \geq 0$, on flat spacetime with metric,
\begin{align}
    \dd s^2=\eta_{\mu \nu} \dd x^\mu \dd x^\nu 
    = -\dd t^2 + \sum_{i=1}^{d-1} \dd x_i^2 \,,
\end{align}
where the signature of $\eta_{\mu \nu}$ is mostly plus, and the indices $\mu, \nu$ run $0$ to $d-1$.
The bulk AdS metric with a radius $L$ in Poincar\'{e} coordinate is 
\begin{align}\label{BCFT_coord}
	\dd s^2 = G_{MN} \dd X^M \dd X^N =L^2 \frac{\dd z^2+\eta_{\mu \nu} \dd x^\mu \dd x^\nu}{z^2} \,,
\end{align}
where $z$ is a radial coordinate and its range is $0<z<\infty$.
To reduce the isometry of the metric from $SO (2,d)$ to $SO (1,d)$, we introduce a boundary $\CQ$ in the radial direction.
To construct the gravity dual following Takayanagi's proposal \cite{Takayanagi:2011zk,Fujita:2011fp}, we introduce the boundary with a brane of tension $T$,
\begin{align} \label{Action}
	\begin{aligned}
	I &=  \frac{1}{16\pi G_\text{N}} \int_{\CB} \! \dd^{d+1}X \, \sqrt{-G}\, \left( R + \frac{d(d-1)}{L^2}\right) \\
    	&\qquad + \frac{1}{8\pi G_\text{N}} \int_{\CQ} \! \dd^{d}X \, \sqrt{-\hat{G}} \left( K - T\right) + \frac{1}{8\pi G_\text{N}} \int_{\CM} \! \dd^d X \, \sqrt{-\hat{G}}\, K \,,
\end{aligned}
\end{align}
where $\CB$ is the bulk AdS spacetime and $\CM$ is the boundary on which the dual BCFT lives.
In the present case, $\CM$ is the half plane, $\CB$ is the bulk AdS spacetime in the coordinates \eqref{BCFT_coord} with the restricted range $x_1 \geq -\alpha z $, and $\CQ$ is the AdS boundary at $x_1= -\alpha z $.
$\hat{G}_{MN}$ represents the  induced metric.
To make the variational problem well-define in the presence of the boundary, the Gibbons-Hawking term is introduced with the extrinsic curvature defined by 
\begin{align}
	K_{MN} = \hat{G}_{ML}\hat{G}_{NK}\nabla^L n^K \,,
\end{align}
for the outward pointing normal vector $n^M$.
The Dirichlet boundary condition is imposed on $\CM$, but the Neumann boundary condition is chosen on $\CQ$
\begin{align}
	K_{MN} -\hat{G}_{MN} K = - T\, \hat{G}_{MN} \,.
\end{align}
Since the extrinsic curvature is given by
\begin{align}
	K = - \frac{d}{L} \frac{x_1}{\sqrt{z^2+x_1^2}} \,,
\end{align}
the brane tension is fixed to be 
\begin{align}
	T = \frac{d-1}{L} \frac{\alpha}{\sqrt{1+\alpha^2}} \,.
\end{align}
For $d=2$, comparing the model to the Liouville setup, we find the parameter relations  $L=1/\sqrt{\mu}$ and $T = \mu_{\mathrm{B}}$ as discussed in the previous section via the boundary entropies. Because $\alpha$ plays the same role in each setup, we do not mind the duplicate notation of $\alpha$.

\subsection{CV conjecture}

Let us compute a holographic complexity following the CV conjecture \eqref{CVcon}
\begin{align}
    C_{\text{V}}= \frac{V}{G_{\text{N}} L} \,,
\end{align} 
in the AdS/BCFT model. 
In this setup, the length scale $L$ is fixed to be the AdS radius and $V$ is the maximum volume at $t=0$ given by
\begin{align}
    V&=\int _\epsilon ^\infty \dd z \int _{-\alpha z}^{\infty} \dd x_1 \int \prod_{i=2}^{d-1} \dd x_i \frac{L^d}{z^d} \\
    &= \frac{1}{2}V_{d-1}L^d \int _\epsilon ^\infty \frac{\dd z}{z^d}
    + \alpha \frac{L^{d}V_{d-2}}{(d-2)\epsilon^{d-2}} \,, \label{integration}
\end{align}
where $\epsilon$ is a cutoff, and $V_{d-1}$ and $V_{d-2}$ are $(d-1)$- and $(d-2)$-dimensional infinite volumes, respectively.
The first term corresponds to a half of a complexity without boundary.
In the CV conjecture \eqref{CVcon}, the boundary complexity \eqref{bdyC} is 
\begin{align}
    \Delta C_{\text{V}}^{\text{bdy}} 
    = C_{\text{V}}^{\text{BCFT}} - \frac{1}{2}C_{\text{V}}^{\text{CFT}}
    = \alpha  \frac{L^{d-1}V_{d-2}}{(d-2)G_{\text{N}}\epsilon^{d-2}} \,. 
\end{align}
The boundary contribution still survives and is proportional to $1/\epsilon^{d-2}$ as expected.
In $d=2$ case, the boundary complexity is logarithmically divergent,
\begin{align}
    \Delta C_{\text{V}}^{\text{bdy}}=\alpha  \frac{L}{G_{\text{N}}} \log \left( \frac{z_\infty}{\epsilon}\right) =
    \frac{2c}{3} \alpha \log \left( \frac{z_\infty}{\epsilon}\right) \,, \label{CVd2}
\end{align}
where $z_\infty$ is an IR cutoff and the relation $c=3L/2G_{\text{N}}$ is used.
This is obtained by a direct computation of the integral \eqref{integration} or a replacement of $V_{d-2}/(d-2)\epsilon^{d-2}$ with $\log (z_\infty /\epsilon)$. 
This result quantitatively matches with that of the path-integral complexity \eqref{bdyCL}.
Since both of the boundary complexity and the boundary entropy are monotonic increasing functions of the slope $\alpha$, they are monotonically decreasing under the boundary RG flow.
Note that the RG flow from UV to IR corresponds to from large $\alpha$ to small $\alpha$.
See \cite{Flory:2017ftd} for a related previous work on the CV conjecture with boundary or defect.

\subsection{CA conjecture}

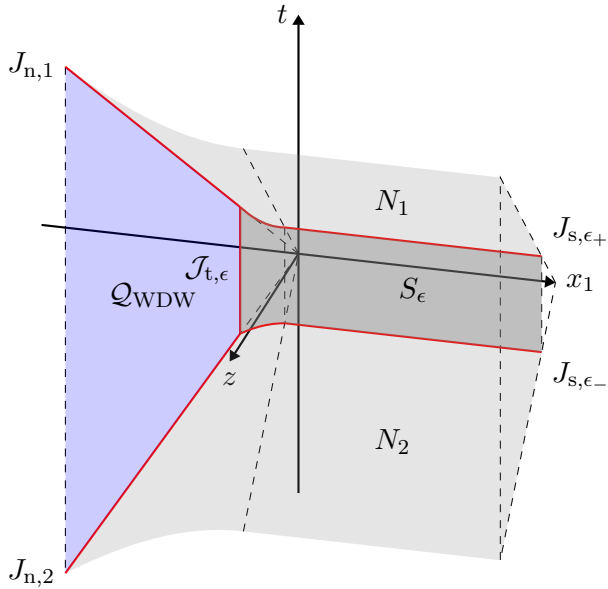
\begin{figure}
\centering
\tdplotsetmaincoords{65}{105} 
\begin{tikzpicture}[tdplot_main_coords,scale=0.7]
%%%coordinate system%%%
\draw[thick,-{Triangle[angle'=60,scale=0.8]}] (0,0,0) -- (5,0,0) node[below]{$z$}; 
\draw[thick,-{Triangle[angle'=60,scale=0.8]}] (0,-5,0) -- (0,5,0) node[right]{$x_1$}; 
\draw[thick,-{Triangle[angle'=60,scale=0.8]}] (0,0,-5) -- (0,0,5) node[left]{$t$};
\coordinate (O) at (0,0,0);
\coordinate (Ob) at (0,5,0);
%%%1st layer z=1%%%
\coordinate (A11) at (1,-{sqrt(3)}/2,{sqrt(1+3/4)});
\coordinate (B11) at (1,0,1);
\coordinate (C11) at (1,5,1);
\coordinate (A12) at (1,-{sqrt(3)}/2,-{sqrt(1+3/4)});
\coordinate (B12) at (1,0,-1);
\coordinate (C12) at (1,5,-1);
%%%2nd layer z=4%%%
\coordinate (A21) at (4,-{4*sqrt(3)}/2,{4*sqrt(1+3/4)});
\coordinate (B21) at (4,0,4);
\coordinate (C21) at (4,5,4);
\coordinate (A22) at (4,-{4*sqrt(3)}/2,-{4*sqrt(1+3/4)});
\coordinate (B22) at (4,0,-4);
\coordinate (C22) at (4,5,-4);
%%%lines%%%
\draw[dashed] (O)--(B11)--(B21) (B22)--(B12)--(O)
             (Ob)--(C11)--(C21)--(C22)--(C12)--cycle 
             (B11)--(B12) (C11)--(C12) 
             (O)--(A11) (O)--(A12) (A21)--(A22);
\draw[red, thick] (A11)--(A12) (A11)--(A21) (A12)--(A22) 
                        (B11)--(C11) (C12)--(B12);
\draw[red,thick,name path=P11] (B11) parabola (A11);
\draw[red,thick,name path=P12] (B12) parabola (A12);
%%%Fillings%%%
\fill[gray,opacity=0.5] (B11)--(C11)--(C12)--(B12)--cycle;
\fill[gray,opacity=0.2] (B11)--(C11)--(C21)--(B21)
                       (B12)--(C12)--(C22)--(B22);
\fill[blue,opacity=0.2] (A11)--(A12)--(A22)--(A21)--cycle;
\fill[domain=0:1,smooth,variable=\t, gray, opacity=0.5] plot(1,{-sqrt(3)/2*\t},{sqrt((sqrt(3)/2*\t)^2+1)})--plot(1,{-sqrt(3)/2*(1-\t)},{-sqrt((sqrt(3)/2*(1-\t))^2+1)})--cycle ;
\fill[domain=0:1,smooth,variable=\t, gray, opacity=0.2]
plot(1,{-sqrt(3)/2*\t},{sqrt((sqrt(3)/2*\t)^2+1)})--plot(4,{-4*sqrt(3)/2*(1-\t)},{4*sqrt((sqrt(3)/2*(1-\t))^2+1)})--cycle 
plot(1,{-sqrt(3)/2*\t},{-sqrt((sqrt(3)/2*\t)^2+1)})--plot(4,{-4*sqrt(3)/2*(1-\t)},{-4*sqrt((sqrt(3)/2*(1-\t))^2+1)})--cycle;
%%%labels%%%
\draw ($(B11)!.5!(C12)$) node {$S_\epsilon$} 
(2.5,-{2.5/2*sqrt(3)},0) node {$\CQ_{\text{WDW}}$}
($(B11)!.5!(C21)$) node {$N_{1}$} ($(B12)!.5!(C22)$) node {$N_{2}$}
(A21) node[left] {$J_{\text{n},1}$}  (A22) node[left] {$J_{\text{n},2}$}
(C11) node[above right] {$J_{\text{s},\epsilon_+}$} (C12) node[below right] {$J_{\text{s},\epsilon_-}$}
($(A11)!.5!(A12)$) node[left] {$\CJ_{\text{t},\epsilon}$};
\end{tikzpicture}
\caption{The WDW patch which is the causal development of the Cauchy slice $t = 0$. The bulk region $\CB_{\text{WDW}}$ is surrounded by a portion $\CQ_{\text{WDW}}$ of the brane $\CQ$, null surfaces $N_{1,2}$ and the timelike surface $S_\epsilon$ at $z=\epsilon$. The red lines are joints $J_{\text{n},1,2} = N_{1,2} \cap \CQ_{\text{WDW}}$, $J_{\text{s},\epsilon_\pm} =  N_{1,2} \cap S_\epsilon$ and $\CJ_{\text{t},\epsilon} = \CQ_{\text{WDW}} \cap S_\epsilon$. The other timelike surface $S_\infty$ at $z=z_\infty$, the other spacelike joints $J_{\text{s},\infty_\pm}$ and the other timelike joint $\CJ_{\text{t},\infty}$ are not depicted.}
\label{fig:HC}
\end{figure}

The CA conjecture \eqref{CAcon} argues that the holographic complexity is given by the WDW action 
\begin{align}
    C_{\text{A}} =\frac{I_{\text{WDW}}}{\pi} \,,
\end{align}
as noted in introduction.
We consider a state at $t=0$, and hence the causal development of the Cauchy slice, called the WDW patch, is surrounded by the boundary $\CQ$, two null surfaces emanating from $(z,t)=(0,0)$, denoted by $N_1$ for the future directing surface and $N_2$ for the past directing surface.
For regularization, we introduce two timelike surfaces at $z=\epsilon$ and $z=z_\infty$, denoted by $S_\epsilon$ and $S_{\infty}$ respectively.
The WDW patch for $x_1\geq 0$ region is the same as that of a pure AdS spacetime.
However, as noted in \cite{Chapman:2018bqj}, the WDW patch for $x_1<0$ region is surrounded by null rays emanating from the point $(z,t,x_1)=(0,0,0)$ and is given by $t^2<z^2+x_1^2$.\footnote{We would like to thank S.~Chapman, D.~Ge and G.~Policastro for pointing out this.}
The WDW patch contains two null joints, $J_{\text{n},1}$ and $J_{\text{n},2}$, four spacelike joints, $J_{\text{s},\epsilon_+}$, $J_{\text{s},\epsilon_-}$, $J_{\text{s},\infty_+}$ and $J_{\text{s},\infty_-}$, and two timelike joints, $\CJ_{\text{t},\epsilon}$ and $\CJ_{\text{t},\infty}$.
See Figure \ref{fig:HC} for the configuration of the WDW patch.
The WDW action consists of variable terms,
\begin{align} 
     	\begin{aligned}
	   I_{\text{WDW}} &=  \frac{1}{16\pi G_\text{N}} \int_{\CB_{\text{WDW}}} \! \dd^{d+1}X \, \sqrt{-G}\, \left( R + \frac{d(d-1)}{L^2}\right) \\
    	&\quad + \frac{1}{8\pi G_\text{N}} \int_{\CQ_{\text{WDW}}} \! \dd^d X \,  \sqrt{-\hat{G}} \left( K - T\right) + \frac{1}{8\pi G_\text{N}} \sum_{i=\epsilon,\infty} \int_{S_i} \! \dd^d X \, \sqrt{-\hat{G}}\, K \\
    	&\quad + \frac{1}{8\pi G_\text{N}} \sum_{i=1}^2 \epsilon_\kappa \left(  \int_{N_i} \! \dd \lambda \dd^{d-1} \bm{x}\, \sqrt{\gamma} \kappa + \int_{N_i} \! \dd \lambda \dd^{d-1} \bm{x} \, \sqrt{\gamma} \Theta \log (\ell_{\text{ct}}|\Theta | ) \right) \\
    	& \quad + \frac{1}{8\pi G_\text{N}} \sum_{J} \epsilon_a \int_{J} \! \dd^{d-1} X \, \sqrt{h} a
    	+\frac{1}{8\pi G_\text{N}} \sum_{\CJ}\epsilon_\phi \int_{\CJ} \! \dd^{d-1} X \, \sqrt{-h} \phi \,. \label{WDWA}
\end{aligned}
\end{align}
The first term is a bulk contribution in the WDW patch, 
$\CB_{\text{WDW}}$ which is a bulk AdS region surrounded by $\CQ$, $N_{1,2}$ and $S_{\epsilon,\infty}$. The second term is the Gibbons-Hawking term with the brane tension $T$ of the boundary region $\CQ$ surrounded by $N_{1,2}$ and $S$, denoted by $\CQ_{\text{WDW}}$.
The third term represents the Gibbons-Hawking term of the cutoff surfaces $S_\epsilon$ and $S_\infty$.
The terms in the third line are null surface contributions and their counter terms, which are introduced for a reparametrization invariance. 
$\epsilon_\kappa= -1$ for future of the boundary segment and $\epsilon_\kappa=1$ for past of the boundary segment. 
$\gamma_{MN}$ is the induced metric on the null surfaces and $\gamma$ is its determinant.
$\kappa$ is defined by the equation $k^M \nabla_M k_N =\kappa k_N$ and represents how the null coordinate $\lambda$ deviates from affine parametrization.
$\Theta=\partial_\lambda \log \sqrt{\gamma} $ represents the expansion.
The new length scale in the counter term, $\ell_{\text{ct}}$, serves a scale appearing in the definition of complexity in quantum field theory.
The rest terms are joint contributions and the details are explained when we evaluate them.
See \cite{Lehner:2016vdi,Hayward:1993my} for the detail of the various terms.

Comments on cutoffs are in order. As usual, we have to introduce the cutoffs at $z=\epsilon$ for a UV reguralization and at $z=\infty$ for an IR reguralization.
Note that a different reguralization scheme is often used in literature about the AdS/CFT setup with defect. See, e.g., \cite{Chapman:2018bqj} for detail.  
In higher dimensions, the IR contributions in \eqref{WDWA} can be ignored compared with other contributions containing UV divergences.

\subsubsection*{Contribution from $\CB_{\text{WDW}}$}

Let us evaluate the bulk action in the WDW patch.
The bulk region in the WDW patch consists of the $x_1 \geq 0$ region $\CB_{\text{WDW}}^+$ and the $x_1 < 0$ region $\CB_{\text{WDW}}^-$.
Since the Ricci scalar of AdS$_{d+1}$ is $R=-d(d+1)/L^2$ , the bulk contribution becomes 
\begin{align}
   I_{\CB_{\text{WDW}}}
    &=\frac{1}{16\pi G_\text{N}} \int_{\CB_{\text{WDW}}} \! \dd^{d+1}X \, \sqrt{-G}\, \left( R + \frac{d(d-1)}{L^2}\right) \\ &=I_{\CB_\text{WDW}^+} -\frac{dL^{d-1}V_{d-2}}{8\pi G_\text{N}(d-2)\epsilon^{d-2}} \left( \alpha \sqrt{1+\alpha^2} + \mathrm{arcsinh}\, \alpha \right) \,.
\end{align}
The first term is a contribution from $\CB_{\text{WDW}}^+$ and a half of the pure AdS spacetime. 
The second term is a contribution by the boundary and comes from the region $\CB_{\text{WDW}}^-$.

\subsubsection*{Contribution from $\mathcal{Q}_{\text{WDW}}$}

Since null rays on the surface are given by $t=\pm \sqrt{1+\alpha^2}z$, the WDW patch on the brane $\CQ_{\text{WDW}}$ is surrounded by the null rays.
The induced metric on the brane is 
\begin{align}
    \dd s^2 = L^2 \frac{(1+\alpha^2)\dd z^2 -\dd t^2 + \sum_{i=2}^{d-1}\dd x_i^2 }{z^2} \,,
\end{align}
and the extrinsic curvature becomes 
\begin{align}
    K = \frac{d}{L}\frac{\alpha}{\sqrt{1+\alpha^2}} \,.
\end{align}
The WDW action of the brane $\CQ$ becomes 
\begin{align}
   I_{\CQ_{\text{WDW}}} 
    &=\frac{1}{8\pi G_\text{N}} \int_{\CQ_{\text{WDW}}} \! \dd^{d}X  \, \sqrt{-\hat{G}} \left( K - T\right) \\
    &= \frac{L^{d-1}V_{d-2}}{4\pi G_\text{N}(d-2)\epsilon^{d-2}}\alpha \sqrt{1+\alpha^2} \,.
 \end{align}
Only the boundary contribution survives.

\subsubsection*{Contribution from $S_\epsilon$ and $S_\infty$}
The induced metric on $S_{\epsilon}$ is
\begin{align}
    \dd s^2=\frac{L^2}{\epsilon^2}\eta_{\mu \nu}\dd x^\mu \dd x^\nu \,,
\end{align}
and the extrinsic curvature on $S_\epsilon$ is 
\begin{align}
    K=\frac{d-1}{L} \,.
\end{align}
Then, the surface contribution on $S_\epsilon$ becomes
\begin{align}
    I_{S_\epsilon}
    &=\frac{1}{8\pi G_{\text{N}}}\int_{S_\epsilon} \! \dd^d X \, \sqrt{-\hat{G}}K \\
    &=I_{S_{\epsilon}^+}
    +\frac{(d-1)L^{d-1}V_{d-2}}{8\pi G_{\text{N}}\epsilon^{d-2}}\left( \alpha\sqrt{1+\alpha^2} + \mathrm{arcsinh} \, \alpha \right) \,,
\end{align}
where the first term is half of the action of the surface at $z=\epsilon$ in the AdS spacetime and the second term is a boundary contribution coming from the $x_1<0$ region.

The contribution for $S_\infty$ can be easily obtained by changing the sign of the extrinsic curvature and replacing $\epsilon$ with $z_\infty$,
\begin{align}
    I_{S_\infty}=I_{S_{\infty}^+}
    -\frac{(d-1)L^{d-1}V_{d-2}}{8\pi G_{\text{N}}z_\infty^{d-2}}\left( \alpha\sqrt{1+\alpha^2} + \mathrm{arcsinh} \, \alpha \right) \,.
\end{align}
For higher dimensions, the IR surface contribution can be ignored while it still survives for $d=2$.

In $d=2$ case, the boundary contributions are opposite and the sum of two boundary contributions vanishes.

\subsubsection*{Contribution from null surfaces}
Since the geometry is symmetric at $t=0$, contributions of two null surfaces are the same.
The null surface consists of the $x_1 \geq 0$ region, $N_1^+$, and the $x_1 < 0$ region, $N_1^-$.
Since we are especially interested in the boundary complexity, we need not to evaluate the WDW action of $N_1^+$. Here we give a brief prescription to evaluate it.
The null surface $N_1^+$ is parameterized by the coordinate $\lambda =z/N=t/N$ where $N$ is an arbitrary parameter, 
\begin{align}
    x^M = (N\lambda, N\lambda, x_1,\bm{x}) \,,
\end{align}
where the first component is $z$-direction and the second component is $t$-direction.
The tangent vector to $N_1$ is 
\begin{align}
    k^M =\frac{\dd x^M}{\dd \lambda} =N(1,1, 0,\bm{0}) \,.
\end{align}
Then the induced metric, $\kappa$ and the expansion on $N_1^+$ are given by
\begin{align}
    \dd s^2=L^2\frac{\delta_{ij}\dd x^i \dd x^j}{z^2} \,, \qquad 
    \kappa= -\frac{2}{\lambda} \,, \qquad 
    \Theta =-\frac{d-1}{\lambda} \,,
\end{align}
respectively, where $i$ and $j$ run in space directions on the boundary of AdS.
By using these, it is possible to obtain the null surface contribution of $N_1^+$.

Next, let us consider the $x_1<0$ region, $N_1^-$.
The null surface $N_1^-$ can be parameterized by 
\begin{align}
    x^M = (M\lambda \cos \theta, M\lambda, - M\lambda \sin \theta,\bm{x}) \,,
\end{align}
where $M$ is an arbitrary parameter.
The tangent vector to $N_1^-$ is 
\begin{align}
    k^M =\frac{\dd x^M}{\dd \lambda} =M(\cos \theta ,1, -\sin \theta ,\bm{0}) \,.
\end{align}
Then the induced metric, $\kappa$ and the expansion on $N_1^-$ are given by
\begin{align}
    \dd s^2=L^2\left(\frac{\dd \theta^2}{\cos ^2 \theta} + \frac{\dd \bm{x}^2}{z^2} \right) \,, \qquad 
    \kappa= -\frac{2}{\lambda} \,, \qquad 
    \Theta =-\frac{d-2}{\lambda} \,,
\end{align}
respectively. 

The contribution from the null boundaries is evaluated as 
\begin{align}
    I_{N_1}=I_{N_2}&=-\frac{1}{8\pi G_\text{N}} \int_{N_1} \! \dd \lambda \dd \theta \dd^{d-2} \bm{x} \, \sqrt{\gamma} \kappa \\
    &=I_{N_1^+} + \frac{L^{d-1}V_{d-2}}{4\pi G_\text{N}(d-2) \epsilon^{d-2}}\mathrm{arcsinh}\, \alpha \,,
\end{align}
and the counter terms are 
\begin{align}
    I_{N_1,\text{ct}}&=I_{N_2,\text{ct}}=-\frac{1}{8\pi G_\text{N}} \int_{N_1} \! \dd \lambda \dd \theta \dd^{d-2}\bm{x} \, \sqrt{\gamma} \Theta \log (\ell_{\text{ct}}|\Theta | )  \\
    &\begin{aligned}
    &=I_{N_1^+,\text{ct}}-\frac{L^{d-1}V_{d-2}}{8\pi G_\text{N} \epsilon^{d-2}}
     \left(\frac{1}{d-2} - \log \left( \frac{\ell_{\text{ct}}M(d-2)}{\epsilon} \right)  \right) \, \mathrm{arcsinh} \alpha \\
    &\qquad \qquad \qquad + \frac{L^{d-1}V_{d-2}}{8\pi G_\text{N} \epsilon^{d-2}} \int_0^{\theta_\alpha} \! \dd \theta \frac{\log \cos \theta}{\cos \theta} \,,
    \end{aligned}
    \end{align}
with $\tan \theta_\alpha = \alpha$.
Here $I_{N_1^+}$ and $I_{N_1,\text{ct}}^+$ represent contributions without boundary.
The boundary contribution of $I_{N_1^+}$ does not depend on the arbitrary parameter $M$, while that of the counter term depends on the arbitrary parameter $M$.
However, this dependence cancels with the joint terms of $J_3$ as we will see later.
Hence, we do not discuss this point anymore here.
Note that the counter term vanishes in $d=2$ since the expansion vanishes, $\Theta =0$.

\subsubsection*{Contribution from spacelike joints}

There are four spacelike joints in the WDW patch.
Two spacelike joints between the null surfaces and $S_\epsilon$ are denoted by $J_{\text{s},\epsilon_+}$ for $t>0$ and $J_{\text{s},\epsilon_-}$ for $t<0$, respectively.
Similarly, there are two spacelike joints between the null surfaces and $S_\infty$ denoted by $J_{\text{s},\infty_+}$ for $t>0$ and $J_{\text{s},\infty_-}$ for $t<0$, respectively.
From the symmetric reason, $J_{\text{s},\epsilon_+}$ and $J_{\text{s},\epsilon_-}$ are the same contribution, and $J_{\text{s},\infty_+}$ and $J_{\text{s},\infty_-}$ are the same contribution. See Fig. \ref{fig:HC}.

Let us compute the joint term $J_{\text{s},\epsilon_+}$, firstly.
The unit normal vector of $S_\epsilon$ is given by
\begin{align}
    s^M = \frac{\epsilon}{L}(-1,0,\cdots,0) \,, 
\end{align}
and a term in the integrand is given by 
\begin{align}
    a=\log |k\cdot s| =\log \left( \frac{ML\cos \theta}{\epsilon} \right) \,.
\end{align}
Then the joint action with $\epsilon_a=-1$ becomes 
\begin{align} \label{Jse}
    I_{J_{\text{s},\epsilon_+}}=I_{J_{\text{s},\epsilon_+}^+} - \frac{L^{d-1}V_{d-2}\mathrm{arcsinh}\, \alpha}{8\pi G_\text{N}\epsilon^{d-2}} \log \left( \frac{ML}{\epsilon} \right) - \frac{L^{d-1}V_{d-2}}{8\pi G_\text{N}\epsilon^{d-2}} \int_0^{\theta_\alpha} \! \dd \theta \, \frac{\log \cos \theta}{\cos \theta}  \,.
\end{align}
The first term represents half of the joint action without boundary, and the second and third terms are just boundary contributions.

The joint term $J_{\text{s},\infty_+}$ can be easily obtained by replacing $\epsilon$ in \eqref{Jse} with $z_\infty$ and $\epsilon_a = -1$ with $\epsilon_a = 1$. For $d>2$, this term can be ignored, but for $d=2$, the contribution remains.

For $d>2$, by adding $I_{N_1,\text{ct}}$ and $I_{J_{\text{s},\epsilon_+}}$, we can easily confirm that the combination between $I_{N_1,\text{ct}}$ and $I_{J_{\text{s},\epsilon_+}}$ does not depend on the arbitrary parameters $N$ and $M$ as mentioned above.
In $d=2$, the sum of $I_{N_1}$, $I_{J_{\text{s},\epsilon_+}}$ and $I_{J_{\text{s},\epsilon_-}}$ does not depend on the arbitrary parameter $M$.

\subsubsection*{Contribution from null joints}

There are two null joints  in the WDW patch.
A null joint between the null surface $N_1$ ($N_2$) and the brane $\CQ$ is denoted by $J_{\text{n},1}$ ($J_{\text{n},2}$).
From the symmetric reason, $J_{\text{n},1}$ and $J_{\text{n},2}$ are the same contribution.

Let us evaluate a contribution of the null joints $J_{\text{n},1}$ and $J_{\text{n},2}$.
The unit normal vector to $\CQ$ is 
\begin{align}
    s^M = \frac{z}{L\sqrt{1+\alpha^2}}(-\alpha,0,-1,0,\cdots,0) \,, 
\end{align}
and the induced metric on the joint is given by
\begin{align}
 \dd s^2 = L^2\frac{\alpha^2 \dd z^2 + \sum_{i=2}^{d-1}\dd x_i^2}{z^2} \,.
\end{align}
Since the vector $s^M$ and the null vector $k^M$ are orthogonal, the integrand of the joint $J_{\text{n},1}$ contains a strong divergence, 
\begin{align}
    a=\log |k\cdot s| =\log 0 \,.
\end{align}
Naively, it seems that this $\log 0$ divergence causes an incurable problem.
However, the joint term $J_{\text{n},1}$ does not depend on the boundary parameter $\alpha$ and it will be subtracted when we define the boundary complexity.

\subsubsection*{Contribution from timelike joints}

The last contribution comes from the timelike joints between the brane $\CQ$ and cutoff surfaces,denoted by $\CJ_{\text{t},\epsilon}$ located at $z=\epsilon$ and $\CJ_{\text{t},\infty}$ located at $z=z_\infty$.
Let us evaluate $\CJ_{\text{t},\epsilon}$, first.
The outgoing normal unit vector to $S_\epsilon$ is 
\begin{align}
   n_{S_\epsilon}^M=\frac{\epsilon}{L} \left( -1,0,\cdots , 0\right) \,, 
\end{align}
and the outgoing normal vector to $\CQ$ is 
\begin{align}
    n_\CQ^M = \frac{\epsilon}{L \sqrt{1+\alpha^2}} \left(-\alpha ,0,-1 ,0,\cdots ,0\right) \,.
\end{align}
Then, the angle between two normal vectors is given by
\begin{align}
    \cos \phi = n_S\cdot n_\CQ =  \frac{\alpha}{\sqrt{1+\alpha^2}} \,.
\end{align}
The joint action is evaluated as 
\begin{align}
    I_{\CJ_{\text{t},\epsilon}}&=\frac{1}{8\pi G_{\text{N}}}\int \! \dd^{d-1}X \, \sqrt{-h}\phi \\ &=\frac{L^{d-1}V_{d-2}}{4\pi G_{\text{N}}\epsilon^{d-2}}\sqrt{1+\alpha^2} \arccos \left( \frac{\alpha}{\sqrt{1+\alpha^2}}\right) \,, 
\end{align}
where $h_{MN}$ is an induced metric on the joint.
See \cite{Hayward:1993my} for the detail of the timelike joint contribution.

Next, let us consider the $\CJ_{\text{t},\infty}$ contribution.
The contribution for higher dimensional case can be ignored and we restrict our attention to a two dimensional case.
For $\CJ_{\text{t},\infty}$, the angle between two normal vectors is given by
\begin{align}
    \cos \phi = - \frac{\alpha}{\sqrt{1+\alpha^2}} \,,
\end{align}
and the joint contribution becomes
\begin{align}
    I_{\CJ_{\text{t},\infty}}=\frac{L}{4\pi G_{\text{N}}}\sqrt{1+\alpha^2} \left( \pi - \arccos \left( \frac{\alpha}{\sqrt{1+\alpha^2}}\right) \right) \,.
\end{align}

\subsubsection*{Total CA}
In total, the boundary complexity for $d > 2$ is given by
\begin{align}\label{CAddim}
\begin{aligned}
    \Delta C_{\text{A}}^{\text{bdy}}&= \frac{L^{d-1}V_{d-2}}{8\pi^2 G_\text{N}\epsilon^{d-2}}\left[ (d-2) \left( \alpha \sqrt{1+\alpha^2}+ \mathrm{arcsinh}\, \alpha \right) 
    +2\log \left(\frac{\ell_{\text{ct}}(d-2)}{L}\right)\mathrm{arcsinh}\, \alpha  \right]  \\
    &\quad +\frac{L^{d-1}V_{d-2}}{4\pi^2 G_\text{N}\epsilon^{d-2}}  \left(\sqrt{1+\alpha^2} \arccos \left( \frac{\alpha}{\sqrt{1+\alpha^2}}\right) -\frac{\pi}{2}\right) \,. 
    \end{aligned} 
\end{align}
and the boundary complexity for $d=2$ is given by
\begin{align}
    \Delta C_{\text{A}}^{\text{bdy}}= \frac{L}{4\pi G_\text{N}}  \left(\sqrt{1+\alpha^2}  - 1 \right) \,. \label{bdyCAd2}
\end{align}
When we subtract the complexity without boundary, we include null joint terms and timelike joint terms to the half of the complexity, $C_{\text{A}}^{\text{CFT}}/2$, such that the boundary complexity vanishes for $\alpha =0$.

It is clear that the boundary complexity \eqref{CAddim} in the CA conjecture does not vanish for $d>2$.
In $d=2$ case, it turns out that the finite term of the boundary complexity \eqref{bdyCAd2} is universal, since the logarithmic divergent term vanishes,
\begin{align}\label{CAun}
   \left. \Delta C_{\text{A}}^{\text{bdy}}\right|_{\log} 
    = 0 \, , \qquad 
    \left. \Delta C_{\text{A}}^{\text{bdy}}\right|_{\mathrm{univ}}
    = \frac{L}{4\pi G_\text{N}}  \left(\sqrt{1+\alpha^2}  - 1 \right) \,.
\end{align}
It means that, for $d=2$, the CA conjecture gives the divergence structure different from the CV conjecture \eqref{CVd2} and the path-integral optimization approach \eqref{bdyCL} which are logarithmically divergent. 

Note that, as $\alpha$ decreases, the boundary complexity \eqref{bdyCAd2} monotonically decreases as same as the boundary entropy \eqref{bdyEE}.

\section{Discussions}
We studied complexity of quantum states in BCFT$_2$ using the path-integral optimization \cite{Caputa:2017urj,Caputa:2017yrh}. 
Since the path-integral optimization naturally produces the AdS geometry with a cutoff in the radial direction as in the AdS/BCFT model \cite{Takayanagi:2011zk,Fujita:2011fp}, we also studied holographic complexity in the AdS$_{d+1}$/BCFT$_d$ model following the CV and the CA conjectures \cite{Susskind:2014,Susskind:2014rva,Brown:2015bva,Brown:2015lvg}. 
It was revealed that the boundary complexity which is an increment of the complexity due to the boundary does not vanish in the path-integral optimization, in the CV conjecture and even in the CA conjecture.
The path-integral complexity and the CV complexity shows logarithmic divergences in $d=2$ case and they are the same up to the overall prefactors.
On the other hand, the CA complexity does not show a logarithmic divergence and has non-vanishing constant in $d=2$.
For higher dimensional case, the CV and the CA boundary complexities show the same divergent structures.

Let us compare our result with Chapman \textit{et al}. \cite{Chapman:2018bqj}.
While the increments of the circuit complexity in several DCFT models vanish in their work, the path-integral complexity increases due to the boundary for a positive $\mu_{\text{B}}$.
Hence this fact implies that whether the boundary complexity and the defect complexity vanish depends on the definition of the complexity in QFT or models in BCFT and DCFT.
In gravity side, our results of the boundary complexity in the CV and the CA conjectures in $d=2$, are consistent with their argument\footnote{In \cite{Chapman:2018bqj} the defect complexity in the CA conjecture vanishes but it was argued  that the coefficient of the logarithmic term vanishes while the finite term depends on the cut-off one employs.
We obtained the non-vanishing contribution in the boundary complexity and argue that it does not depend on the cut-off scheme and it is universal in this sense.} 
and we arrive at the same conclusion which the boundary or defect can distinguish action from volume. 
On the other hand, in higher dimensional case ($d>2$), the boundary complexity does not vanish even in the CA conjecture.
We conclude that the boundary and the defect can not detect the definite difference of the CV conjecture and the CA conjecture except a special case in contrast to the argument by Chapman \textit{et al}. \cite{Chapman:2018bqj}.
We can infer the reason why the boundary complexity in the CA conjecture in the AdS$_3$/CFT$_2$ setup from our higher dimensional calculations.  
The boundary contribution among the volume $\CB_{\text{WDW}}$, the brane $\CQ_{\text{WDW}}$ and the null surfaces $N$ is proportional to $d-2$ and this proportional factor delete a factor $1/(d-2)$ in front of $1/\epsilon^{d-2}$. 
Hence $\log (z_\infty / \epsilon)$ terms do not appear in $d \to 2$ limit.

There is a comment on the boundary Liouville action \eqref{LA} used in the path-integral optimization.
The region $\CM$ has the two boundaries $\partial \CM_0$ and $\partial \CM_1$ and there is a point where the two boundaries cross.
The extrinsic curvature suddenly changes at the point and should be proportional to a delta function.
In such a case, the Liouville action can contain a joint contribution as in \cite{Hayward:1993my} and the complexity would change.
However, even if the contribution coming from the point is added, it does not make the complexity vanishing since it should be proportional to the angle between two normal vectors to the boundaries. It might be notable that the corner angle contribution also appeared in the CA complexity \eqref{CAddim} in $d>2$.

Some comments on the CA conjecture are in order.
The WDW patch in AdS spacetime with boundary in the radial direction includes null joint terms between the boundary and the null surfaces like $J_{\text{n},1}$ and $J_{\text{n},2}$ in our setup.
Since the normal vector of the boundary and the null vector of the null surfaces are perpendicular each other, the joint terms contain unavoidable divergences due to $\log 0$.\footnote{We again thank S.~Chapman, D.~Ge and G.~Policastro for pointing out this to us.}
For the boundary complexity, we avoid this problem by including similar terms in the subtracted WDW action of the half AdS spacetime. 
However, the CA complexity with boundary itself suffers from the divergences coming from $\log 0$.
The resolution of this problem might be shed light on the holographic complexity.

To simplify our discussion in the path-integral optimization, we restricted our attention to BCFT$_2$. 
It is interesting to apply the path-integral optimization method to defect CFT or higher-dimensional BCFT and confirm whether the defect or the boundary contributions still survive.
Also, finite temperature variants of the AdS/BCFT models, which the holographic complexities were studied in  \cite{Flory:2017ftd,Cooper:2018cmb,Numasawa:2018grg}, might be good playgrounds.

In two-dimensional CFTs, \cite{Caputa:2018kdj} made connection between the circuit complexity \cite{Jefferson:2017sdb} and the path-integral optimization \cite{Caputa:2017urj,Caputa:2017yrh} 
by using a geometric action associated with Virasoro group (see also \cite{Camargo:2019isp,Akal:2019hxa}).
It is interesting to generalize \cite{Caputa:2018kdj} to BCFT$_2$ and check whether the boundary complexity vanishes or not.
We hope for the non-vanishing boundary complexity in this setup.

\acknowledgments

YS would like to thank A.~O'Bannon for introducing me to the paper \cite{Chapman:2018bqj}.
The authors thank S.~Chapman, D.~Ge, K.~Goto, Y.~Nakayama, S.~Sugishita, T.~Takayanagi and K.~Umemoto for fruitful discussions.
The authors sincerely thank the authors of the paper \cite{Chapman:2018bqj} for pointing out our mistakes in terms of the WDW patch and E.~Tonni for correspondence.
The authors thank the Yukawa Institute for Theoretical Physics at Kyoto University.
Discussions during the workshop YITP-T-19-03 ``Quantum Information and String Theory 2019'' were useful to complete this work.
The work of YS is supported by JSPS KAKENHI Grant-in-Aid (Wakate-A), No.17H04837 and the work of KW is supported by the Grant-in-Aid for Japan Society for the Promotion of Science Fellows, No.18J00322.

\bibliographystyle{JHEP}
\bibliography{ComplexityBCFT}

\providecommand{\href}[2]{#2}\begingroup\raggedright\begin{thebibliography}{10}

\bibitem{Ryu:2006bv}
S.~Ryu and T.~Takayanagi, {\it {Holographic derivation of entanglement entropy
  from AdS/CFT}},  {\em Phys. Rev. Lett.} {\bf 96} (2006) 181602,
  [\href{http://arxiv.org/abs/hep-th/0603001}{{\tt hep-th/0603001}}].

\bibitem{Rangamani:2016dms}
M.~Rangamani and T.~Takayanagi, {\it {Holographic Entanglement Entropy}},  {\em
  Lect. Notes Phys.} {\bf 931} (2017) pp.1--246,
  [\href{http://arxiv.org/abs/1609.01287}{{\tt arXiv:1609.01287}}].

\bibitem{Susskind:2014moa}
L.~Susskind, {\it {Entanglement is not enough}},  {\em Fortsch. Phys.} {\bf 64}
  (2016) 49--71, [\href{http://arxiv.org/abs/1411.0690}{{\tt
  arXiv:1411.0690}}].

\bibitem{Susskind:2014}
L.~Susskind, {\it {Computational Complexity and Black Hole Horizons}},  {\em
  Fortsch. Phys.} {\bf 64} (2016) 24--43,
  [\href{http://arxiv.org/abs/1402.5674}{{\tt arXiv:1402.5674}}].

\bibitem{Susskind:2014rva}
L.~Susskind, {\it {Addendum to Computational Complexity and Black Hole
  Horizons}},  {\em Fortsch. Phys.} {\bf 64} (2016) 44--48,
  [\href{http://arxiv.org/abs/1403.5695}{{\tt arXiv:1403.5695}}].

\bibitem{Brown:2015bva}
A.~R. Brown, D.~A. Roberts, L.~Susskind, B.~Swingle, and Y.~Zhao, {\it
  {Holographic Complexity Equals Bulk Action?}},  {\em Phys. Rev. Lett.} {\bf
  116} (2016), no.~19 191301, [\href{http://arxiv.org/abs/1509.07876}{{\tt
  arXiv:1509.07876}}].

\bibitem{Brown:2015lvg}
A.~R. Brown, D.~A. Roberts, L.~Susskind, B.~Swingle, and Y.~Zhao, {\it
  {Complexity, action, and black holes}},  {\em Phys. Rev.} {\bf D93} (2016),
  no.~8 086006, [\href{http://arxiv.org/abs/1512.04993}{{\tt
  arXiv:1512.04993}}].

\bibitem{MIyaji:2015mia}
M.~Miyaji, T.~Numasawa, N.~Shiba, T.~Takayanagi, and K.~Watanabe, {\it
  {Distance between Quantum States and Gauge-Gravity Duality}},  {\em Phys.
  Rev. Lett.} {\bf 115} (2015), no.~26 261602,
  [\href{http://arxiv.org/abs/1507.07555}{{\tt arXiv:1507.07555}}].

\bibitem{Belin:2018fxe}
A.~Belin, A.~Lewkowycz, and G.~Sárosi, {\it {The boundary dual of the bulk
  symplectic form}},  {\em Phys. Lett.} {\bf B789} (2019) 71--75,
  [\href{http://arxiv.org/abs/1806.10144}{{\tt arXiv:1806.10144}}].

\bibitem{Belin:2018bpg}
A.~Belin, A.~Lewkowycz, and G.~Sárosi, {\it {Complexity and the bulk volume, a
  new York time story}},  {\em JHEP} {\bf 03} (2019) 044,
  [\href{http://arxiv.org/abs/1811.03097}{{\tt arXiv:1811.03097}}].

\bibitem{Caputa:2017urj}
P.~Caputa, N.~Kundu, M.~Miyaji, T.~Takayanagi, and K.~Watanabe, {\it {Anti-de
  Sitter Space from Optimization of Path Integrals in Conformal Field
  Theories}},  {\em Phys. Rev. Lett.} {\bf 119} (2017), no.~7 071602,
  [\href{http://arxiv.org/abs/1703.00456}{{\tt arXiv:1703.00456}}].

\bibitem{Caputa:2017yrh}
P.~Caputa, N.~Kundu, M.~Miyaji, T.~Takayanagi, and K.~Watanabe, {\it {Liouville
  Action as Path-Integral Complexity: From Continuous Tensor Networks to
  AdS/CFT}},  {\em JHEP} {\bf 11} (2017) 097,
  [\href{http://arxiv.org/abs/1706.07056}{{\tt arXiv:1706.07056}}].

\bibitem{Jefferson:2017sdb}
R.~Jefferson and R.~C. Myers, {\it {Circuit complexity in quantum field
  theory}},  {\em JHEP} {\bf 10} (2017) 107,
  [\href{http://arxiv.org/abs/1707.08570}{{\tt arXiv:1707.08570}}].

\bibitem{Camargo:2019isp}
H.~A. Camargo, M.~P. Heller, R.~Jefferson, and J.~Knaute, {\it {Path integral
  optimization as circuit complexity}},  {\em Phys. Rev. Lett.} {\bf 123}
  (2019), no.~1 011601, [\href{http://arxiv.org/abs/1904.02713}{{\tt
  arXiv:1904.02713}}].

\bibitem{Bhattacharyya:2018wym}
A.~Bhattacharyya, P.~Caputa, S.~R. Das, N.~Kundu, M.~Miyaji, and T.~Takayanagi,
  {\it {Path-Integral Complexity for Perturbed CFTs}},  {\em JHEP} {\bf 07}
  (2018) 086, [\href{http://arxiv.org/abs/1804.01999}{{\tt arXiv:1804.01999}}].

\bibitem{Takayanagi:2018pml}
T.~Takayanagi, {\it {Holographic Spacetimes as Quantum Circuits of
  Path-Integrations}},  {\em JHEP} {\bf 12} (2018) 048,
  [\href{http://arxiv.org/abs/1808.09072}{{\tt arXiv:1808.09072}}].

\bibitem{Chapman:2017rqy}
S.~Chapman, M.~P. Heller, H.~Marrochio, and F.~Pastawski, {\it {Toward a
  Definition of Complexity for Quantum Field Theory States}},  {\em Phys. Rev.
  Lett.} {\bf 120} (2018), no.~12 121602,
  [\href{http://arxiv.org/abs/1707.08582}{{\tt arXiv:1707.08582}}].

\bibitem{Caputa:2018kdj}
P.~Caputa and J.~M. Magan, {\it {Quantum Computation as Gravity}},  {\em Phys.
  Rev. Lett.} {\bf 122} (2019), no.~23 231302,
  [\href{http://arxiv.org/abs/1807.04422}{{\tt arXiv:1807.04422}}].

\bibitem{Chapman:2018bqj}
S.~Chapman, D.~Ge, and G.~Policastro, {\it {Holographic Complexity for Defects
  Distinguishes Action from Volume}},  {\em JHEP} {\bf 05} (2019) 049,
  [\href{http://arxiv.org/abs/1811.12549}{{\tt arXiv:1811.12549}}].

\bibitem{Azeyanagi:2007qj}
T.~Azeyanagi, A.~Karch, T.~Takayanagi, and E.~G. Thompson, {\it {Holographic
  calculation of boundary entropy}},  {\em JHEP} {\bf 03} (2008) 054--054,
  [\href{http://arxiv.org/abs/0712.1850}{{\tt arXiv:0712.1850}}].

\bibitem{Takayanagi:2011zk}
T.~Takayanagi, {\it {Holographic Dual of BCFT}},  {\em Phys. Rev. Lett.} {\bf
  107} (2011) 101602, [\href{http://arxiv.org/abs/1105.5165}{{\tt
  arXiv:1105.5165}}].

\bibitem{Fujita:2011fp}
M.~Fujita, T.~Takayanagi, and E.~Tonni, {\it {Aspects of AdS/BCFT}},  {\em
  JHEP} {\bf 11} (2011) 043, [\href{http://arxiv.org/abs/1108.5152}{{\tt
  arXiv:1108.5152}}].

\bibitem{BCT}
P.~Braccia, A.~Cotrone, and E.~Tonni.
\newblock \textit{In preparation}.

\bibitem{Miyaji:2016mxg}
M.~Miyaji, T.~Takayanagi, and K.~Watanabe, {\it {From path integrals to tensor
  networks for the AdS/CFT correspondence}},  {\em Phys. Rev.} {\bf D95}
  (2017), no.~6 066004, [\href{http://arxiv.org/abs/1609.04645}{{\tt
  arXiv:1609.04645}}].

\bibitem{Fateev:2000ik}
V.~Fateev, A.~B. Zamolodchikov, and A.~B. Zamolodchikov, {\it {Boundary
  Liouville field theory. 1. Boundary state and boundary two point function}},
  \href{http://arxiv.org/abs/hep-th/0001012}{{\tt hep-th/0001012}}.

\bibitem{Flory:2017ftd}
M.~Flory, {\it {A complexity/fidelity susceptibility $g$-theorem for
  AdS$_{3}$/BCFT$_{2}$}},  {\em JHEP} {\bf 06} (2017) 131,
  [\href{http://arxiv.org/abs/1702.06386}{{\tt arXiv:1702.06386}}].

\bibitem{Affleck:1991tk}
I.~Affleck and A.~W.~W. Ludwig, {\it {Universal noninteger 'ground state
  degeneracy' in critical quantum systems}},  {\em Phys. Rev. Lett.} {\bf 67}
  (1991) 161--164.

\bibitem{Friedan:2003yc}
D.~Friedan and A.~Konechny, {\it {On the boundary entropy of one-dimensional
  quantum systems at low temperature}},  {\em Phys. Rev. Lett.} {\bf 93} (2004)
  030402, [\href{http://arxiv.org/abs/hep-th/0312197}{{\tt hep-th/0312197}}].

\bibitem{Casini:2016fgb}
H.~Casini, I.~S. Landea, and G.~Torroba, {\it {The g-theorem and quantum
  information theory}},  {\em JHEP} {\bf 10} (2016) 140,
  [\href{http://arxiv.org/abs/1607.00390}{{\tt arXiv:1607.00390}}].

\bibitem{Lehner:2016vdi}
L.~Lehner, R.~C. Myers, E.~Poisson, and R.~D. Sorkin, {\it {Gravitational
  action with null boundaries}},  {\em Phys. Rev.} {\bf D94} (2016), no.~8
  084046, [\href{http://arxiv.org/abs/1609.00207}{{\tt arXiv:1609.00207}}].

\bibitem{Hayward:1993my}
G.~Hayward, {\it {Gravitational action for space-times with nonsmooth
  boundaries}},  {\em Phys. Rev.} {\bf D47} (1993) 3275--3280.

\bibitem{Cooper:2018cmb}
S.~Cooper, M.~Rozali, B.~Swingle, M.~Van~Raamsdonk, C.~Waddell, and D.~Wakeham,
  {\it {Black Hole Microstate Cosmology}},  {\em JHEP} {\bf 07} (2019) 065,
  [\href{http://arxiv.org/abs/1810.10601}{{\tt arXiv:1810.10601}}].

\bibitem{Numasawa:2018grg}
T.~Numasawa, {\it {Holographic Complexity for disentangled states}},
  \href{http://arxiv.org/abs/1811.03597}{{\tt arXiv:1811.03597}}.

\bibitem{Akal:2019hxa}
I.~Akal, {\it {Reflections on Virasoro circuit complexity and Berry phase}},
  \href{http://arxiv.org/abs/1908.08514}{{\tt arXiv:1908.08514}}.

\end{thebibliography}\endgroup

\end{document}